\documentstyle[graphicx,aps,multicol]{revtex}
\draft
\begin{document}
\title{Level densities and $\gamma$-strength functions in $^{148,149}$Sm}
\author{S. Siem\thanks{Electronic address: sunniva.siem@fys.uio.no}
, M. Guttormsen, K.~Ingeberg, E. Melby, J. Rekstad and A. Schiller\footnote{Present address: Lawrence Livermore National Laboratory, L-414, 7000 East Avenue, Livermore CA 94551}\\
Department of Physics, University of Oslo,
P.O.Box 1048, Blindern, N-0316 Oslo, Norway\\
A. Voinov\\
Frank Laboratory of Neutron Physics, Joint Institute of Nuclear Research, 141980 Dubna, Moscow reg., Russia}
\maketitle
\begin{abstract}
The level densities and $\gamma$-strength functions of the weakly deformed $^{148}$Sm and $^{149}$Sm nuclei have been extracted. The temperature versus excitation energy curve, derived within the framework of the micro canonical ensemble, shows structures, which we associate with the break up of Cooper pairs. The nuclear heat capacity is deduced within the framework of both the micro canonical and the canonical ensemble. We observe negative heat capacity in the micro canonical ensemble whereas the canonical heat capacity exhibits an S-shape as function of temperature, both signals of a phase transition. The structures in the $\gamma$-strength functions are discussed in terms of the pygmy resonance and the scissors mode built on exited states. The samarium results are compared with data for the well deformed $^{161,162}$Dy, $^{166,167}$Er and $^{171,172}$Yb isotopes and with data from (n,$\gamma$)-experiments and giant dipole resonance studies.
\end{abstract}

\pacs{PACS number(s): 21.10.Ma, 24.10.Pa, 27.60+j, 24.30.Gd}
\begin{multicols}{2}

\section{Introduction}
Nuclear structure changes significantly when approaching closed shells. The change of the nuclear shape from deformed to spherical is expected to influence the level density as well as the $\gamma$-strength function, which are important in order to calculate cross-sections\cite{hauser} and (n,$\gamma$) reaction rates\cite{Go96}. These quantities are critical input parameters in astrophysical models describing the nucleosynthesis in stars\cite{rauscher}.

The Oslo Cyclotron group has developed a method to simultaneously extract the level density and $\gamma$-strength function, over a wide excitation energy and $\gamma$-energy region, from measured $\gamma$-ray spectra \cite{Henden95,17}. Today in the rare earth region only mid-shell nuclei have been investigated\cite{MB99,critisk,voinov,elin01}. Therefore a natural extension of these studies is to pin down the properties for nuclei close to the $N = 82$ gap.
This work presents experimental data on level densities and $\gamma$-strength functions of the weakly deformed $^{148}$Sm and $^{149}$Sm nuclei.
The present experimental technique 
represents the least model-dependent method to obtain the $\gamma$ strength function over a wide energy region below the neutron-separation energy.
It is actually the density of levels accessible to the nuclear system, in the $\gamma$-decay process, which is extracted experimentally and interpreted as the level density. This interpretation is approximately valid for a thermalized nucleus in the continuum.
The main advantage of utilizing $\gamma$-rays as a probe for level density is that the nuclear system is likely to be thermalized prior to the $\gamma$-emission. 

A long-standing problem in experimental nuclear physics has been to observe the transition from a super-fluid state, at the ground state, to a normal (Fermi gas) state at higher temperature. The nuclear level density can be utilized to deduce e.g.~entropy, temperature and heat capacity and these observables can reveal indications of a phase transition. 

The $\gamma$-strength function is a measure for the average electromagnetic properties of nuclei and has a fundamental importance for the understanding of nuclear structure and reactions involving $\gamma$ rays~\cite{kop}. At $\sim 3$ MeV of $\gamma$ energy, a bump is observed in the $\gamma$-strength function of rare earth nuclei from ($^3$He,$\alpha$) experiments~\cite{voinov}. The structure in the $\gamma$-ray strength
function, usually called pygmy resonance, will be discussed in terms of the scissors mode built on exited states. The samarium results are compared to results for well deformed $^{161,162}$Dy, $^{166,167}$Er and $^{171,172}$Yb isotopes, with data from (n,$\gamma$)-experiments~\cite{iga-sm99} and giant dipole resonance data.
 
Section~II describes the experimental methods. In Sect.~III the experimental level density and $\gamma$-strength function of  $^{148}$Sm  and $^{149}$Sm are obtained. Section~IV examines thermodynamic properties within the micro-canonical and the canonical ensemble, while electromagnetic properties of the two nuclei are discussed in Sect.~V. Conclusions are made in Sect.~VI.

\section{Method}
The experiment was carried out at the Oslo Cyclotron Laboratory(OCL). 
The neutron pick-up ($^3$He,$\alpha$) and the inelastic scattering
($^3$He,$^3$He')-reactions where employed 
on a $^{149}$Sm self-supporting target, with the 45~MeV $^3$He-beam 
delivered by the MC-35 cyclotron. 
In the inelastic scattering reaction collective excitations are probably more pronounced than in the pick-up reaction, however, in Ref.~\cite{elastisk} it is shown that the two reactions give similar results.

Charged particles and  $\gamma$-rays were 
recorded with the multi-detector system CACTUS\cite{GA90}, 
which contains 8 Si particle telescopes and 27 NaI $\gamma$-ray detectors.
The particle telescopes are placed at an angle of 45$^{\circ}$ relative to 
the beam axis, and consist of a Si front ($\Delta$E) and a Si(Li) back (E) 
detector with thickness $140$ and $3000$ {\normalfont $\mu$}m, respectively.
The NaI $\gamma$-detector array, having a resolution of $\sim 6$ \% at 
$\gamma$ energy $E_{\gamma} = 1$ MeV and a total efficiency of $\sim 15$ \%, 
surrounds the target and particle detectors.
In addition, two Ge detectors were used to monitor the spin distribution and 
selectivity of the reactions. The typical spin window in the two reactions employed is 2-6 $\hbar$.

\begin{figure}[htb]
  \centering
  \includegraphics[angle=-0,width=7.0in]
                {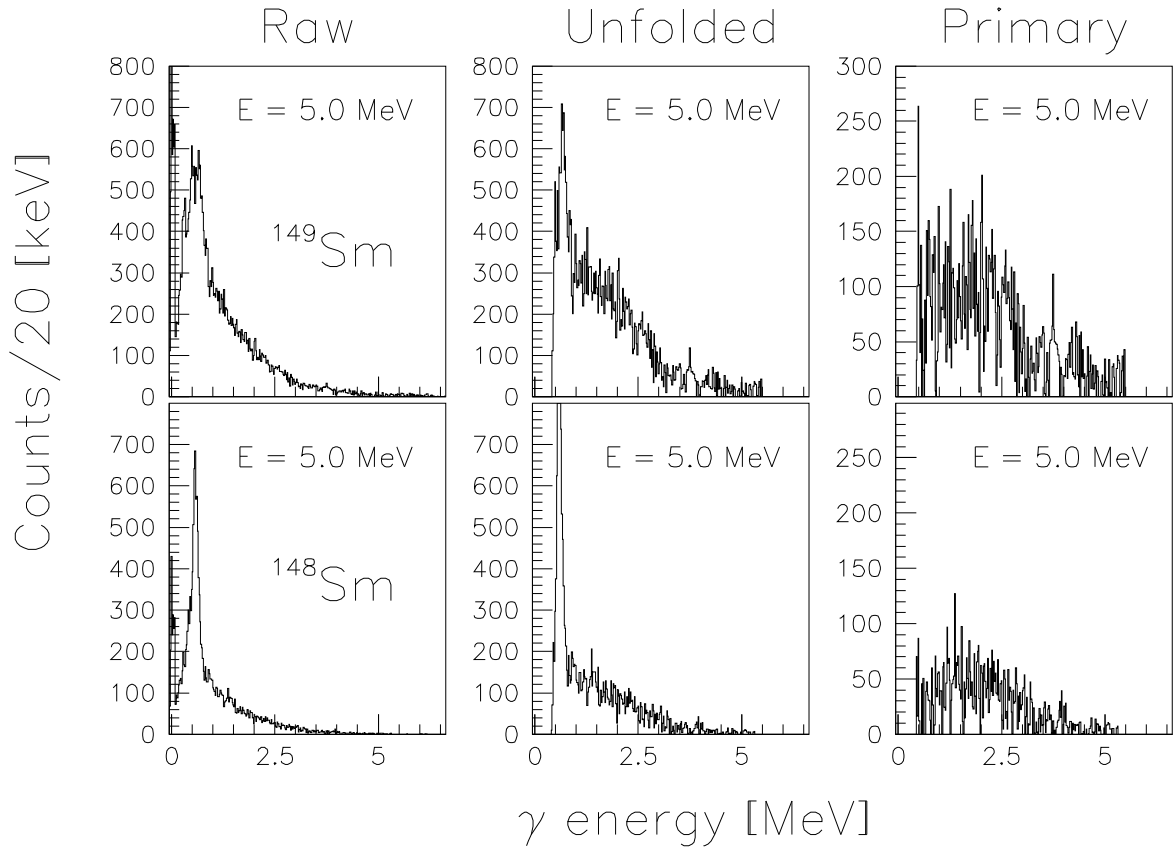}
\caption{Gamma-ray spectra at 5 MeV of excitation energy for $^{149}$Sm (top) and $^{148}$Sm (bottom). The raw (left), unfolded (middle) and primary (right) $\gamma$-ray spectra are shown.}
\label{fig:raw}
\end{figure}

From the reaction kinematics the measured ejectile energy can be 
transformed to excitation energy $E$. Thus, each coincident $\gamma$-ray can 
be assigned to a $\gamma$-cascade originating from a specific 
excitation energy. 
The data are sorted into a matrix of $(E,E_{\gamma})$ energy pairs. Examples of the recorded $\gamma$ spectra, the so called raw spectra, from 5 MeV excitation energy are shown in the left panels of Fig.~\ref{fig:raw} for $^{149}$Sm (top) and $^{148}$Sm (bottom). Note that the statistics of $^{149}$Sm is about twice as good as for $^{148}$Sm, since the ($^3$He,$^3$He') reaction has a higher cross section than the ($^3$He,$\alpha$) reaction.
For each excitation energy $E$ the NaI $\gamma$-ray spectra are corrected for the measured response function of the NaI detectors with the unfolding procedure of Ref.~\cite{GT96}. Unfolded $\gamma$ spectra for the two nuclei are shown in the middle panels of Fig.~\ref{fig:raw}.

From the corrected $\gamma$-ray spectra the primary $\gamma$-ray matrix was extracted according to the subtraction technique of Refs.\cite{Henden95,magne87}. Examples of primary $\gamma$ spectra can be seen in the right panels of Fig.~\ref{fig:raw}.
The method of extracting the primary $\gamma$ spectra is based on the assumption that the decay properties of a bin of excited states are independent of whether the states are directly populated through the nuclear reaction or from de-excitation from higher excited states. This is believed to be approximately fulfilled because the long life time of excited states gives the nucleus time to thermalize prior to the $\gamma$ decay.

\section{Extraction of level density and $\gamma$-strength functions}
The energy distribution of primary $\gamma$-rays provides information on both the level density and the $\gamma$-ray strength function, enabling a simultaneous determination of the two functions.
The fundamental assumption behind the extraction procedure is the Brink-Axel hypothesis~\cite{15,Ax62}, where the probability of $\gamma$-decay in the statistical regime, represented by the primary $\gamma$ matrix $P(E, E_{\gamma})$, can be expressed simply as a product of the final-state level density $\rho(E-E_{\gamma})$ and a $\gamma$-energy dependent factor $F(E_{\gamma})$
\begin{equation}
P(E, E_{\gamma}) \propto F(E_{\gamma}) \rho(E-E_{\gamma}).
\label{eq:ab}
\end{equation}
From the $\gamma$-energy dependent factor $F(E_{\gamma})$ the $\gamma$-strength function can be found. The details of the method and the assumptions behind the factorization of this expression are described in Ref.~\cite{17} and only an outline of the normalization procedures is given here.
Equation~(\ref{eq:ab}) has an infinite number of solutions. It can be shown~\cite{17} that all equally good solutions of Eq.~(\ref{eq:ab}) can be obtained by the transformations
\begin{eqnarray}
\tilde{\rho}(E-E_{\gamma}) &=& A \exp[\alpha(E-E_{\gamma})] \rho(E-E_{\gamma}) , 
\label{eq:array}
\\
\tilde{F}(E_{\gamma}) &=& B \exp(\alpha E_{\gamma}) F(E_{\gamma}) 
\label{eq:array2}
\end{eqnarray} 
of any particular solution ($\rho,F$).
Consequently, neither the slope nor the absolute value of the two functions can be obtained through the iteration procedure, but the three variables $A$, $B$ and $\alpha$ of Eqs.~(\ref{eq:array}) and~(\ref{eq:array2}) have to be determined independently to give the best physical solution of the level density and $\gamma$-strength function.

\subsection{Level density}
\begin{figure}[htb]
  \centering
  \includegraphics[angle=-0,width=7.0in]
                {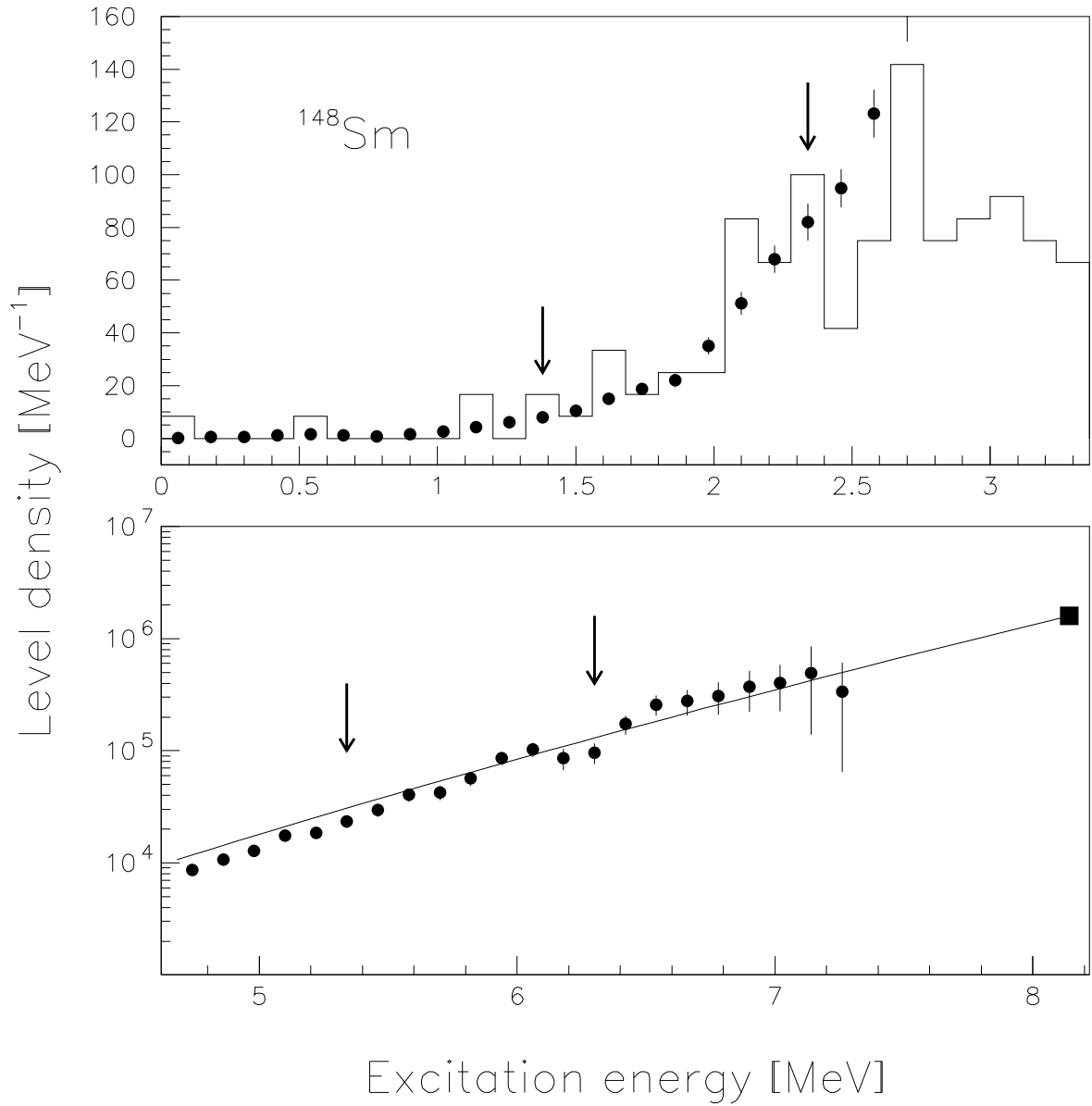}
\caption{The experimental level density for $^{148}$Sm (data points) is 
normalized between the arrows to known levels from spectroscopy \protect\cite{FS96} at low 
excitation energy (histogram upper panel) and to the level density calculated 
from neutron-resonance spacings\protect\cite{IM92} at the neutron separations energy (square data point in lower panel). The solid line is a Fermi-gas approximation.}
\label{fig:levels}
\end{figure}

The parameters $A$ and $\alpha$ of Eq.~(\ref{eq:array}) can be determined by fitting the level density from the iteration procedure\cite{17} to the number of known discrete levels~\cite{FS96} at low excitation energy and to the level density estimated from neutron-resonance spacing data at high excitation energy~\cite{IM92}.
This normalization procedure is shown for $^{148}$Sm in Fig.~\ref{fig:levels}. The deduced level density is fitted to the discrete levels (between the arrows in the upper panel of Fig.~\ref{fig:levels}) as far up in energy as we can assume that all levels are known.
At high excitation energies the deduced level density is fitted (between the arrows in the lower panel of Fig.~\ref{fig:levels}) to a Fermi-gas approximation of the level density (line).The Fermi-gas approximation is forced (multiplied by a constant $f$ given in Table~\ref{tab:par}) to pass through the level-density estimate at the neutron-separation energy (filled square) obtained from the neutron-resonance spacing data.
The level density at the neutron binding energy $S_n$ is estimated using the 
Fermi-gas expression:
\begin{equation} 
\rho_{Sn}= \frac{2\sigma^2}{D}\frac{1}{(I+1)\exp(-(I+1)^2/2\sigma^2) + I\exp(-I^2/2\sigma^2)}
\end{equation}
where $D$ is the neutron resonance spacing, 
$I$ is the spin of the target nucleus in neutron resonance experiments (I($^{148}$Sm) = 0 $\hbar$ and I($^{147}$Sm) = 7/2$\hbar$), the spin-cutoff parameter $\sigma$ is defined by $\sigma^2=0.0888\sqrt{a(S_n-E_{bs})}A^{2/3}$, while $A$ is the mass number, and $E_{bs}$ the back-shift parameter\cite{egidy}.  
Some of the parameters used  are given in Table~\ref{tab:par}. 

The level densities for $^{148}$Sm and $^{149}$Sm as a function of excitation energy are shown in 
Fig.~\ref{fig:2levels}. The fact that the even nuclei has lower level density 
is well understood as due to pairing making the even nuclei more bound by bringing the ground state down in energy. 
The physical interpretation of these data is given in Sect.~IV.

\subsection{Gamma-strength function}
Blatt and Weisskopf~\cite{blatt} suggested to adopt the ratio of the partial radiative width $\Gamma_i(E_{\gamma})$ and the level spacing of the initial states $i$ with equal spin and parity $D_i$ in order to describe the $\gamma$ decay in continuum. According to this, the definition of the $\gamma$-ray strength function is given by $f_{XL}=\Gamma_i(E_{\gamma})/(E_{\gamma} ^{2L+1}D_i)$, where $X$ denotes the electric or magnetic character, and $L$ defines the multipolarity of the $\gamma$ transition. 
After the normalization of the level density, the parameter $\alpha$ of Eq.~(\ref{eq:array2}) is already fixed, 
while the normalization constant $B$ remains to be determined.
The $\gamma$-energy dependent factor $F(E_{\gamma})$ is proportional to 
\begin{equation}
\sum_{XL}E_{\gamma}^{2L + 1}f_{XL}(E_{\gamma})
\end{equation}
where $f_{XL}(E_{\gamma})$ is the $\gamma$-ray strength function for the multipolarity $XL$.
We assume that the $\gamma$ decay in the continuum of nuclei with low spin is mainly governed by electric and magnetic dipole radiation and that the accessible levels have equal numbers of positive and negative parity states. Thus, the observed $F(E_{\gamma})$ can be expressed by a sum of the $E1$ and $M1$ $\gamma$-strength functions only:
\begin{equation}
BF(E_{\gamma})=
[f_{E1}(E_{\gamma})+f_{M1}(E_{\gamma})] E_{\gamma}^3.
\label{eq:prim1}
\end{equation}
The  average total radiative width of neutron resonances  $\langle\Gamma_\gamma(S_n,I,\pi)\rangle$~\cite{kop} with excitation energy equal to the neutron-separation energy $S_n$, spin $I$, and parity $\pi$
\begin{eqnarray}
\langle\Gamma_\gamma(S_n,I,\pi)\rangle=\frac{1}{\rho(S_n, I, \pi)} \sum_{XL}\sum_{I_f, \pi_f}&&
\int_0^{S_n}{\mathrm{d}}E_{\gamma} E_{\gamma}^{2L+1} \nonumber\\
f_{XL}(E_{\gamma})
 \rho(S_n-E_{\gamma}, I_f, \pi_f), &&
\label{eq:norm}
\end{eqnarray}
can be written in terms of $F(E_{\gamma})$ by means of Eq.~(\ref{eq:prim1}) for $L = 1$. 
The experimental value of the average total radiative width $\langle\Gamma_\gamma\rangle$ is the weighted sum of Eq.~(\ref{eq:norm}) with contributions from all different $I$ accessible in the $(n,\gamma)$ reaction.
Thus, with the experimental level density $\rho$ already normalized, the normalization constant $B$ can be deduced. 
The average total radiative width of neutron resonances used for the normalization of the $^{148}$Sm $\gamma$-strength function is $\langle\Gamma_\gamma\rangle= 69(2)$~meV, as given in Ref.~\cite{tecdoc} and for $^{149}$Sm $\langle\Gamma_\gamma\rangle$~=~45 meV from Ref.~\cite{NPA93} is used. 
A detailed description of the normalization of the $\gamma$-strength functions is given in Ref.~\cite{voinov}.
The normalized $\gamma$-strength functions of $^{148}$Sm and 
$^{149}$Sm are shown in Fig.~\ref{fig:2styrke}. The observed strength functions differ from each other with less than a factor of two.

In order to check the quality of the level density and $\gamma$-strength functions, the total $\gamma$-spectra for $^{148}$Sm and $^{149}$Sm (originating from excitation energy equal to the neutron binding energy) has been calculated and compared with experimental ones in Fig.~\ref{fig:iga}. The data points with error bars are taken from the $^{147}$Sm(n,$\gamma$)$^{148}$Sm and $^{148}$Sm(n,$\gamma$)$^{149}$Sm experiments\cite{iga-sm99}. The solid line is calculated from the $\gamma$-strength functions and level densities extracted from the present $^{148}$Sm($^3$He,$\alpha \gamma$)$^{149}$Sm  and $^{148}$Sm($^3$He,$^3$He'$\gamma$)$^{148}$Sm data. The calculation is performed by averaging over 100 keV intervals.
The fact that the agreement it so good gives added confidence that our method is working well.

\section{Thermodynamical nuclear properties\label{sect:thermo}}
The present experimental technique has been previously applied to several well deformed rare earth nuclei. 
In Fig.~\ref{fig:8level} the level density functions for $^{148,149}$Sm are compared with the level density functions for the $^{161,162}$Dy\cite{voinov}, $^{166,167}$Er\cite{elin01} and $^{171,172}$Yb\cite{voinov} nuclei. The level density is less for the samarium nuclei indicating less single particle orbitals in the vicinity of the Fermi-energy. This is understood in terms of shell effects: the samarium nuclei, with $N = 86$ and $87$, are closer to the $N = 82$ neutron gap which means there are fewer orbitals below the neutron Fermi-energy.
 
After establishing the level density as a function of excitation energy, we can explore various thermodynamical properties of the nucleus\cite{MB99,critisk,magne62}. The experimental level density $\rho(E)$ is proportional to the number of states accessible to the nuclear system, in this reaction, at excitation energy $E$. 
Nuclei are isolated systems with well defined energy, so the proper way to describe their statistical properties is to use the micro-canonical ensemble. However, the canonical ensemble, permitting heat exchange, and the grand-canonical ensemble, which in addition allows particle exchange, are often used due to mathematical difficulties with detailed calculations within the micro-canonical ensemble. In this work, thermodynamic properties of the $^{148,149}$Sm nuclei will be extracted and discussed using both the micro-canonical and the canonical ensemble.

\subsection{Micro-canonical ensemble}
\begin{figure}[htb]
  \centering
  \includegraphics[angle=-0,width=7.0in]
                {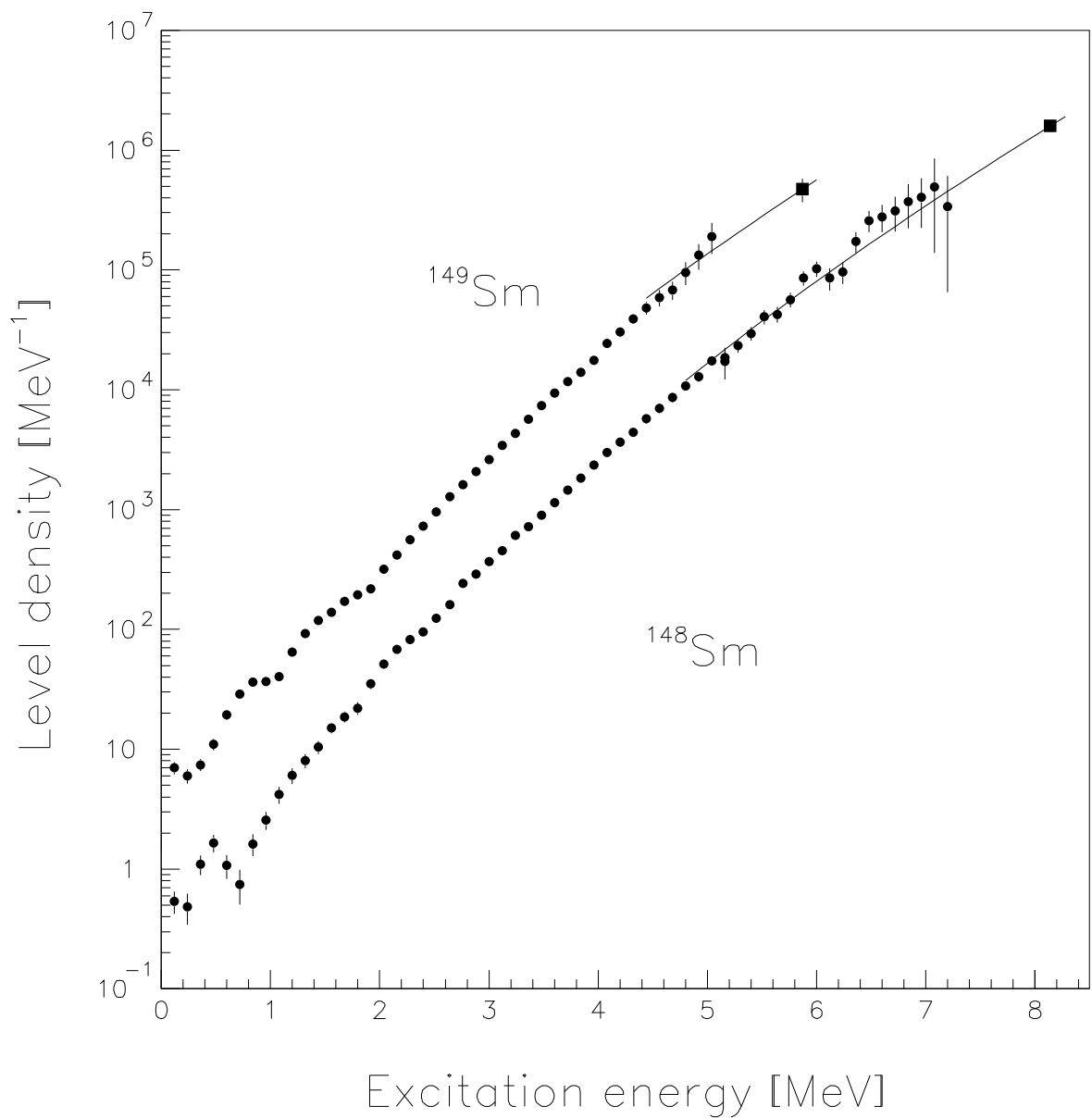}
\caption{The level densities for $^{148}$Sm and 
$^{149}$Sm (data points). The solid lines are  
Fermi gas approximations for the level densities. The filled squares are the level densities at the neutron binding energy calculated 
from neutron-resonance spacings\protect\cite{IM92}.}
\label{fig:2levels}
\end{figure}

\begin{figure}[htb]
  \centering
  \includegraphics[angle=-0,width=7.0in]
                {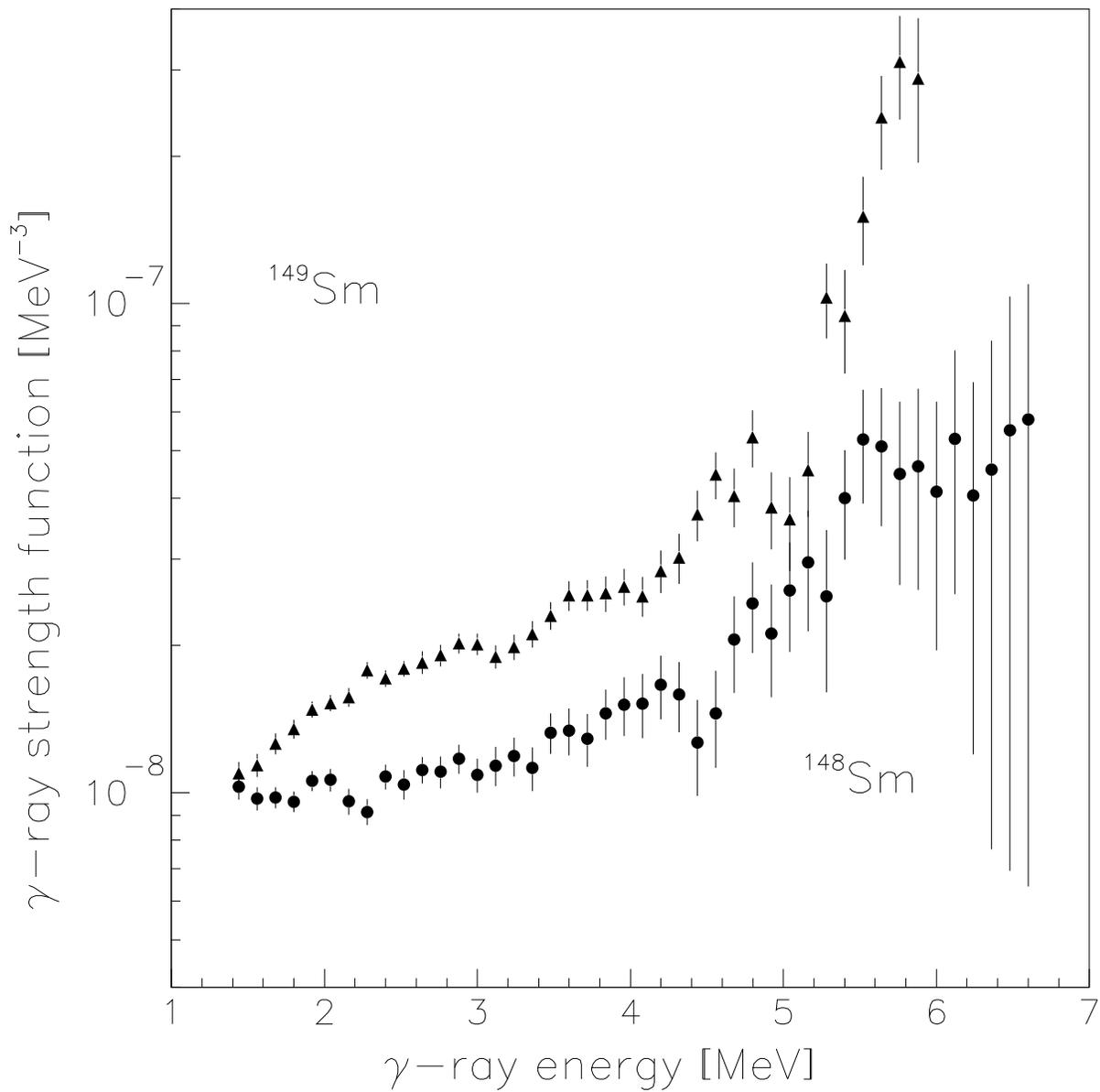}
\caption{The $\gamma$-strength functions for $^{148}$Sm (circles) and 
$^{149}$Sm (triangles). The error bars are due to statistical uncertainties, only.}
\label{fig:2styrke}
\end{figure}

\begin{figure}[htb]
  \centering
  \includegraphics[angle=-0,width=7.0in]
                {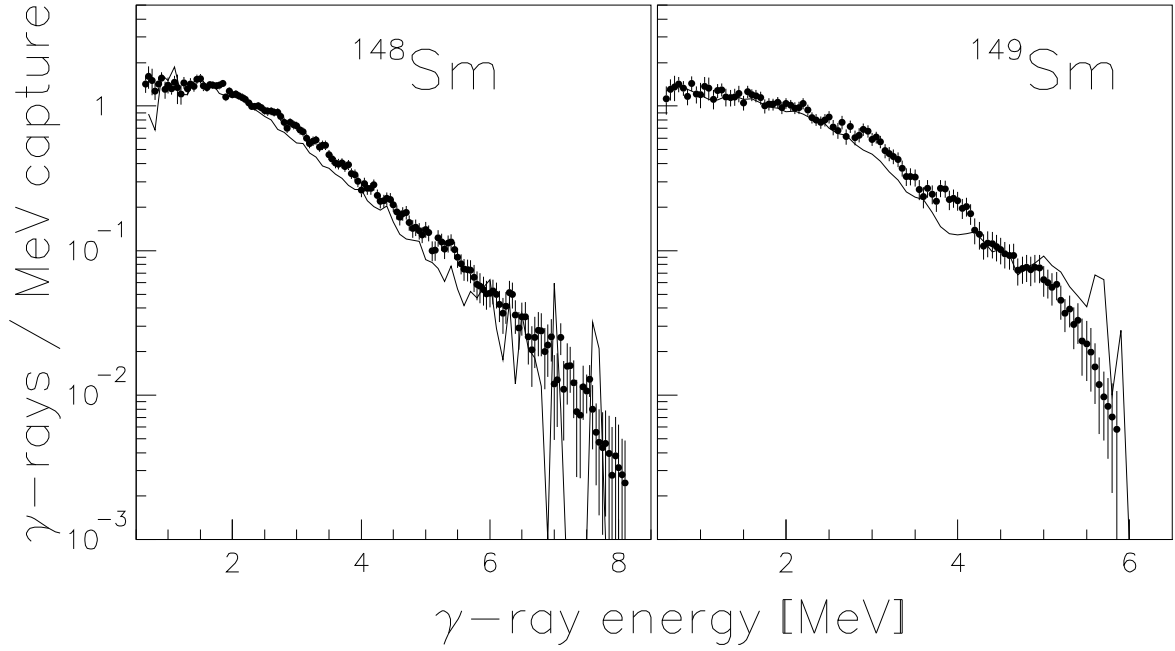}
\caption{The total $\gamma$-spectrum for $^{148}$Sm and $^{149}$Sm measured from excitation energy equal to the neutron binding energy. The data points with error bars are taken from $^{147}$Sm(n,$\gamma$)$^{148}$Sm and $^{148}$Sm(n,$\gamma$)$^{149}$Sm experiments\protect\cite{iga-sm99}. The solid line is calculated from the $\gamma$-strength functions and level densities extracted from the present $^{148}$Sm($^3$He,$\alpha$)$^{149}$Sm  and $^{148}$Sm($^3$He,$^3$He')$^{148}$Sm reaction. The calculation is performed by averaging over 100 keV intervals.}
\label{fig:iga}
\end{figure}

\begin{figure}[htb]
  \centering
\vspace*{2cm}
  \includegraphics[angle=-0,width=7.0in]
                {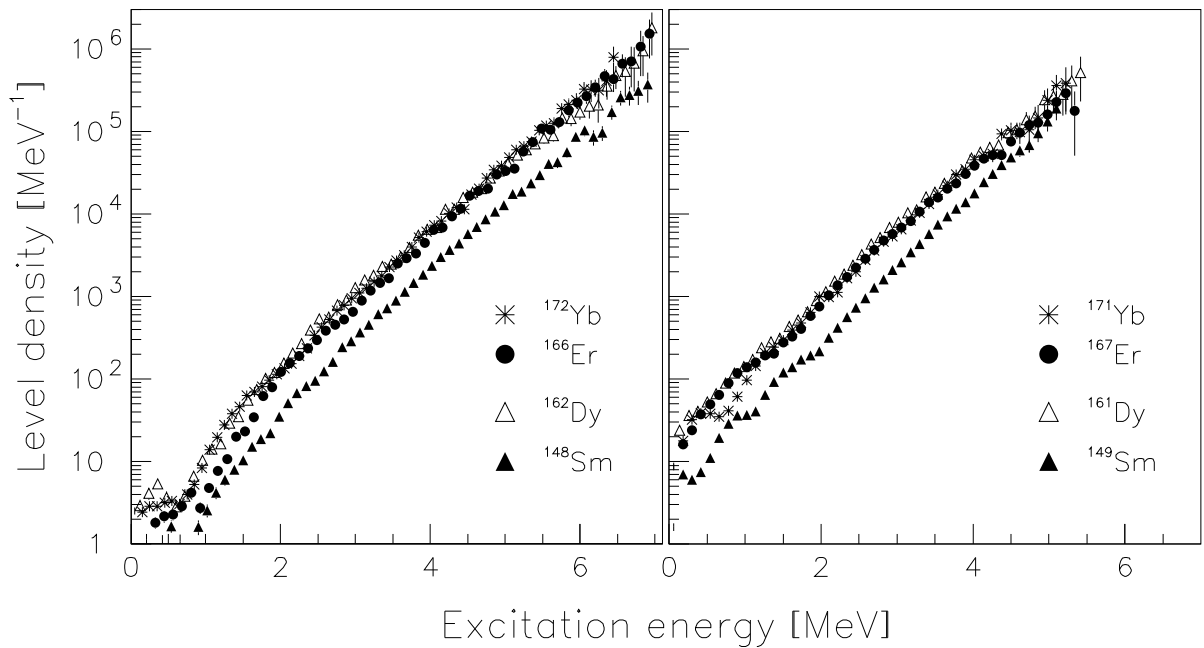}
\caption{The level density functions for $^{148,149}$Sm (filled triangles),
$^{161,162}$Dy\protect\cite{voinov} (open triangles), $^{166,167}$Er\protect\cite{elin01} (filled circles) and $^{171,172}$Yb\protect\cite{voinov} (stars).}
\label{fig:8level}
\end{figure}

When using the micro-canonical ensemble the partition function is simply the multiplicity of nuclear states $\omega (E)$, which experimentally corresponds to the level density of accessible states. Thus, the entropy is defined as:
\begin{equation} S(E)= k_B\ln \omega (E), \end{equation}
where $\omega (E) = \rho(E)/\rho_0$ and $k_B$ is the Boltzmann constant. The normalization constant $\rho_0$ can be adjusted to fulfill the condition of the third law of thermodynamics; $S\rightarrow 0$ when $T\rightarrow 0$, $T$ being the nuclear temperature. The ground-state band of even-even nuclei is assumed to have $T=0$. Therefore, $S_0$ is determined so that the entropy of the ground-state band in $^{148}$Sm is approximately zero. From Fig.~\ref{fig:2levels} we see that $\rho(E=0)=0.3$ MeV$^{-1}$ which determines $\rho_0 = 0.3$ MeV$^{-1}$.  
The entropy curve plotted on a linear scale is almost identical to the level density curve on a logarithmic scale as shown in Fig.~\ref{fig:2levels}. The difference in entropy between the Sm nuclei
\begin{equation}
\delta S(E)=k_B\ln \rho(E,^{149}$Sm$) - k_B\ln \rho(E,^{148}$Sm$)  
\end{equation} seems to be constant and equal $\sim 1.8 k_B$ in the excitation energy region between $E$ = 1.8 - 4.8 MeV. This has also been seen before in the well deformed nuclei \cite{magne62}. As discussed in detail in Ref.~\cite{magne62} $\delta S(E)$ can be understood as the entropy of the unpaired neutron. 

\begin{figure}[htb]
  \centering
  \includegraphics[angle=-0,width=7.0in]
                {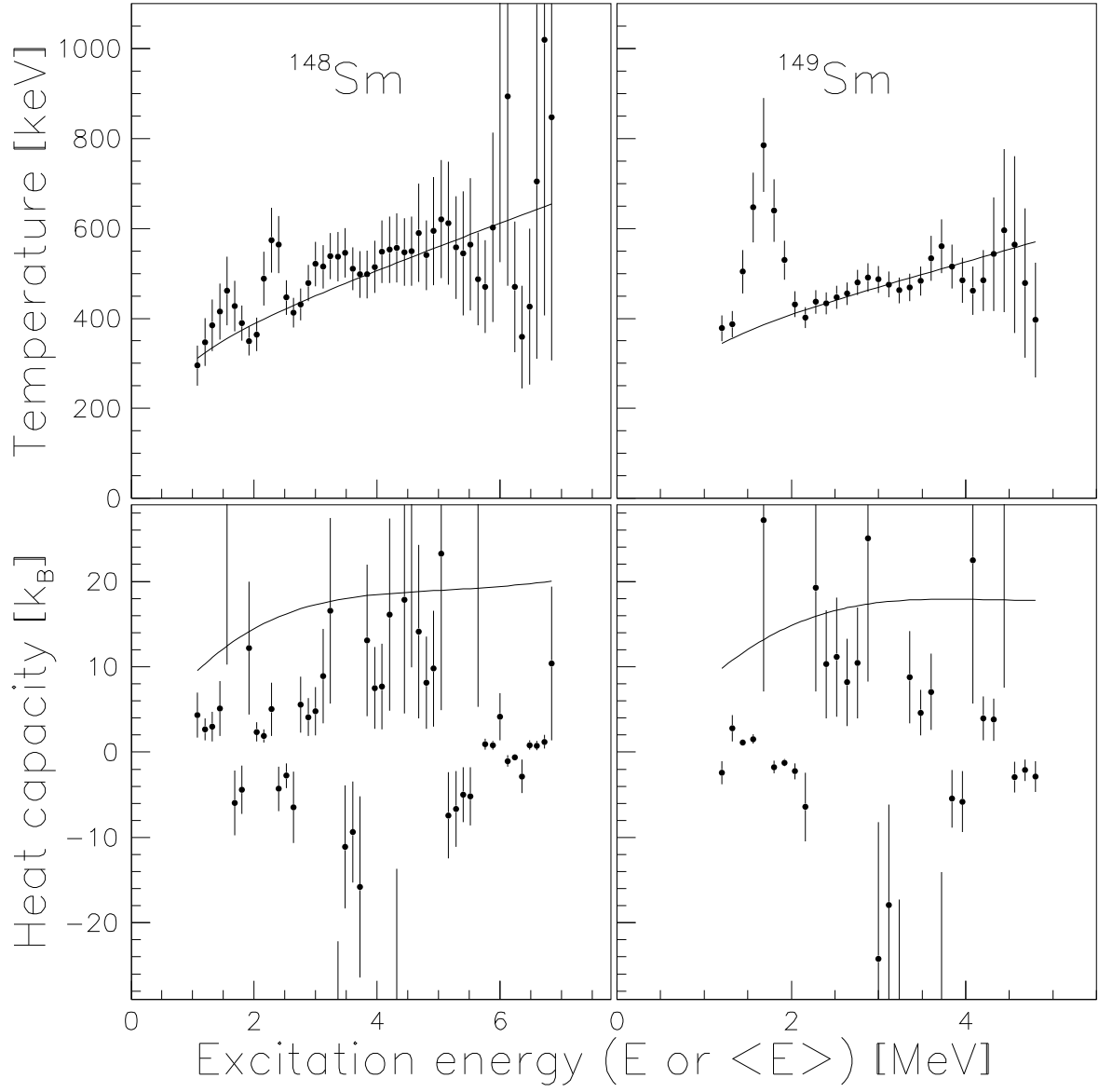}
\caption{Temperature and heat capacity for $^{148}$Sm and $^{149}$Sm derived within the micro canonical ensemble (filled circles) and canonical ensemble (solid lines).}
\label{fig:fig2}
\end{figure}

The micro-canonical temperature in units of MeV is given by:
\begin{equation} T(E)=k_B\left(\frac{\partial S}{\partial E}\right)^{-1}. \label{eq:t}
\end{equation}
The deduced temperature spectra exhibit pronounced bump structures, since the small bumps in the entropy curves are enhanced through the differentiation performed in Eq.~(\ref{eq:t}), as shown for $^{148,149}$Sm in the upper panel of Fig.~\ref{fig:fig2}. The bump structures are interpreted as the breaking of nucleon pairs. 
When particle pairs are broken new degrees of freedom open up i.e.~a more than normal opening of new domains of the phase space, leading to an unusual increase in the entropy and thus the decrease in the temperature. 
We therefore interpret that the location of the break up of a nucleon pair happens in the region where the micro-canonical temperature curve has a negative slope. 
From Fig.~\ref{fig:fig2} we find that this takes place at $E\sim$~1.9 MeV for $^{149}$Sm. This value can be compared to $2\Delta_n$  and $2\Delta_p$, which are the expected cost of breaking up a neutron or proton pair. 
The pairing gap parameters $\Delta_p$ and $\Delta_n$ can be determined from empirical masses of a sequence of isotones or isotopes\cite{bm1}. 
This gives 
2$\Delta_p$ = 2.1~MeV and 2$\Delta_n$ = 2.2~MeV for $^{149}$Sm, which is close to 1.9~MeV found from Fig.~\ref{fig:fig2}. 
For $^{148}$Sm, 
the corresponding values are 2$\Delta_p$ = 2.7~MeV and 2$\Delta_n$~=~2.0~MeV
while the first negative slope in the temperature curve of Fig.~\ref{fig:fig2} is at about 1.75~MeV and the second negative slope is at about 2.5~MeV. 
The first negative slope in the temperature as a function of excitation energy in $^{161}$Dy, $^{162}$Dy, $^{171}$Yb, $^{172}$Yb\cite{critisk} and $^{167}$Er\cite{elin01} also coincides roughly with 2$\Delta$.
While there is a deviation from $2\Delta$ in the localization of the break up of the first pair of particles in the case of $^{166}$Er\cite{elin01}. As discussed in Ref.~\cite{elin01} this can be due to the influence of structural effects in the nuclei, such as e.g.~the Fermi-level position in the Nilsson single-particle scheme, variation in the density of single-particle orbitals, and two quasi-particle couplings to collective degrees of freedom.

The nuclear heat capacity deduced within the framework of the micro canonical ensemble can be determined from differentiating the temperature :
\begin{equation} C_V(E) =k_B\left(\frac{\partial T}{\partial E}\right)^{-1}
\end{equation}
and is shown in Fig.~\ref{fig:fig2} for $^{148}$Sm and $^{149}$Sm. The double differentiation of the entropy introduces strong fluctuations in the heat capacity and the decrease in the micro-canonical temperature lead to the spectacular feature of negative heat capacity. 
The negative heat capacity has been seen before in $^{166}$Er and $^{167}$Er and discussed in Ref.~\cite{elin01}. It has also recently been observed experimentally in the critical region of nuclear fragmentation in Au quasi-projectile sources formed in Au+Au collisions~\cite{agost}. 
There is some controversy on how to interpret negative heat capacities in small systems. Negative heat capacity is usually taken as a signal of a first order phase transition\cite{gross} suggesting that the process of breaking a nucleon pair around $E\sim 2\Delta$ appears like a phase transition of first order. Others claim that negative heat capacity in a small system like the nucleus is compatible with a second order phase transition\cite{Belic}.

\subsection{Canonical ensemble}
The transformation from the micro-canonical to the canonical ensemble is performed by the canonical partition function 
\begin{equation}
Z(T)=\sum_{E=0}^{\infty}\omega (E)e^{-E/T}.
\label{eq:z}
\end{equation}
The partition function is a Laplace transform of the multiplicity of states 
$\omega (E)=\rho (E)/\rho_0$. 
The experimental level density is only covering the excitation energy region from zero to about $S_n-1$ MeV. In the region of and above the neutron-separation energy $S_n$, the Fermi-gas model is believed to describe the nuclear properties. Therefore, the experimental level density is extrapolated to higher excitation energies with the Fermi-gas approximation of Ref.~\cite{egidy}, (see the solid lines of Fig.~\ref{fig:levels}) in order to evaluate Eq.~(\ref{eq:z}).
The thermal average of the excitation energy in the canonical ensemble is
\begin{equation}
\langle E(T)\rangle =Z^{-1}\sum_{E=0}^{\infty}E \omega (E)e^{-E/T}.
\label{eq:e}
\end{equation}
By the Laplace transform in Eq.~(\ref{eq:z}) much of the information contained in the micro-canonical level density becomes smeared out\footnote{The complete information is recovered when using a complex number for the temperature and thus defining the canonical partition function in the complex temperature plane.}. 
The lines in the upper panel of Fig.~\ref{fig:fig2} display the smooth variance of the canonical temperature as a function of the thermal average of the excitation energy.

\begin{figure}[htb]
  \centering
  \includegraphics[angle=-0,width=7.0in]
                {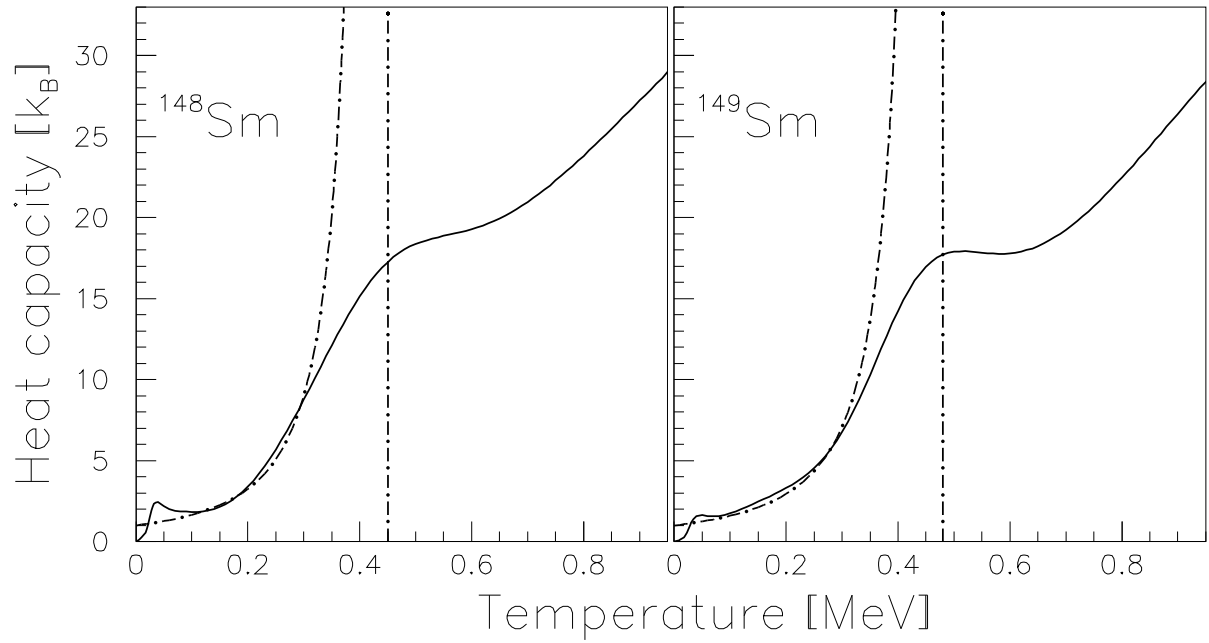}
\caption{The heat capacity for $^{148}$Sm and $^{149}$Sm as a function of temperature derived within the canonical ensemble. The dashed dotted line describes the constant temperature estimate, Eq.~(\ref{eq:cvt}), where $\tau$ is recognized as the critical temperature and marked with the vertical line.}
\label{fig:scurve}
\end{figure}

The nuclear heat capacity deduced within the framework of the canonical ensemble can be used as a "thermometer" for the quenching of pair correlations. It is the derivative of the thermal average of the excitation energy:
\begin{equation}
C_V(T)=k_B\frac{\partial \langle E\rangle }{\partial T}.
\end{equation}
The canonical heat capacities for $^{148}$Sm and $^{149}$Sm are shown in Fig.~\ref{fig:scurve} and exhibit an S-shape as a function of temperature indicating a second order phase transition 
\cite{critisk}. 
The micro-canonical ensemble gives a detailed description of the breaking of one, two, three,... nucleon pairs as a function of excitation energy, while the canonical ensemble reveals the general average properties of this phase-like transition. The proper definition of thermally driven first- and second-order phase transitions in systems with few particles is a long standing problem, which will not be discussed in the present experimental work.

The level density curve of $^{149}$Sm is more linear in a log-plot than the one for $^{148}$Sm (Fig.~\ref{fig:2levels}), this leads to a more pronounced  S-shape in the canonical heat capacity of the odd $^{149}$Sm nuclei compared to the even $^{148}$Sm nuclei, as shown in Fig.~\ref{fig:scurve}. At this point there is a striking difference between the samarium data presented here and earlier data on well deformed mid-shell rare earth nuclei. 
The opposite odd-even effect has been observed experimentally earlier between $^{161}$Dy and $^{162}$Dy, $^{171}$Yb and $^{172}$Yb~\cite{critisk}, and $^{166}$Er and $^{167}$Er\cite{elin01}, where the even nuclei show a more pronounced S-shape than the odd mass isotopes. 
Clearly there are different mechanisms at play here. In the region close to the $N = 82$ close shell the odd nuclei are more deformed than the even nuclei. With deformation more single particle orbitals overlap giving more quasi-particle pairs the chance to interact, which again can be the reason for the sharper rise in the heat capacity curve giving a more pronounced S-shape.
The differences in the S-shape of the heat capacity curves is an interesting observation which awaits further theoretical efforts.

The critical temperature for the quenching of pair correlations can be found by fitting the canonical heat capacity with a constant temperature expression at low excitation energy\cite{critisk}. Assuming a constant temperature level density gives this expression for the canonical heat capacity: 
\begin{equation}C_V(T)=k_B(1-T/\tau)^{-2},
\label{eq:cvt}
\end{equation}
where $\tau$ is the constant-temperature parameter identified with the critical temperature. The dash-dotted lines of Fig.~\ref{fig:scurve} describe Eq.~(\ref{eq:cvt}) with asymptotes at $\tau$ = $T_c$. The critical temperature for the quenching of pair correlations is found at $T_c = 0.45(5)$~MeV for $^{148}$Sm and $T_c = 0.48(5)$~MeV for $^{149}$Sm, in the same range as for the deformed rare earth nuclei \cite{critisk}. The difference in $T_c$ is within the error bars. 

\section{Discussion of the $\gamma$-strength function}
There are several models developed for the $\gamma$-ray strength function $f_{XL}$. We assume that the $\gamma$ decay in the continuum is dominated by dipole transitions. In Ref.~\cite{voinov} various theoretical models for $E1$ and $M1$ radiation were tested and compared to experimental data. In this work only the $E1$ and $M1$ models, which gave the best description of the data, will be used to model the $^{148}$Sm and $^{149}$Sm $\gamma$-strength functions.

The model of Kadmenski{\u{\i}}, Markushev, and Furman (KMF)~\cite{kad} is adopted to account for the $E1$ radiation:
\begin{equation} f_{E1}(E_\gamma)=\frac{1}{3\pi^2\hbar^2c^2} \frac{0.7\sigma_{E1}\Gamma_{E1}^2(E_\gamma^2+4\pi^2T^2)} {E_{E1}(E_\gamma^2-E_{E1}^2)^2}. \end{equation} 
This model~\cite{kad} takes into account the energy and temperature dependence of the GEDR width and is often utilized to describe experimental data~\cite{kop}. 
To be consistent with recent analysis performed for deformed nuclei \cite{voinov}, 
the temperature parameter $T$ is utilized here as a constant fit parameter.
This assumption of a constant temperature is in accordance with the Axel-Brink 
hypothesis, Eq.~(\ref{eq:ab}), giving the same strength function for all excitation energies.
The values for the giant electric dipole resonance parameters $\sigma_{E1}$, $\Gamma_{E1}$ and $E_{E1}$ for $^{148}$Sm are taken from~\cite{GEDR}, for $^{149}$Sm the values are unknown so the ones for $^{148}$Sm were used.  

The $M1$ radiation is described by a Lorentzian based on the existence of a giant magnetic dipole resonance (GMDR)\cite{voinov,kop}:
\begin{equation}
f_{M1}(E_\gamma)=\frac{1}{3\pi^2\hbar^2c^2} \frac{\sigma_{M1}E_\gamma\Gamma_{M1}^2} {(E_\gamma^2-E_{M1}^2)^2+E_\gamma^2\Gamma_{M1}^2}, \label{eq:M1} \end{equation}
where $\sigma_{M1}$, $\Gamma_{M1}$ and $E_{M1}$ are the giant magnetic dipole resonance (GMDR) parameters, which are taken from~\cite{tecdoc}. 
There are experiments which indicate such an $M1$ giant resonance due to spin-flip excitations in the nucleus \cite{kop2}.

\begin{figure}[htb]
  \centering
  \includegraphics[angle=-0,width=7.0in]
                {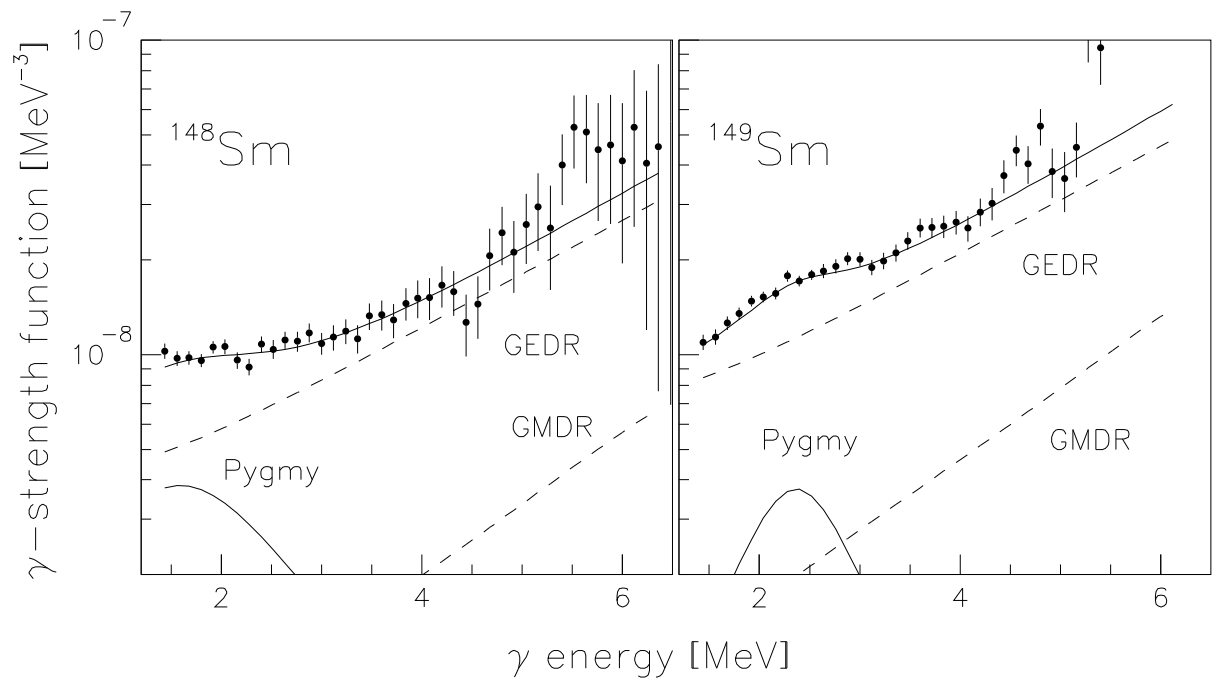}
\caption{The experimental $\gamma$-ray strength functions (data points) of 
$^{148}$Sm (left) and $^{149}$Sm (right). The solid line is the fit to the data 
by the theoretical model. The dashed lines are the respective contributions of the
GEDR $f_{E1}$, the GMDR $f_{M1}$, and the pygmy resonance $f_{\em py}$ to the total theoretical strength function (solid line).}
\label{fig:teori}
\end{figure}

The pygmy resonance is described with a similar Lorentzian function $f_{\rm py}$ as Eq.~(\ref{eq:M1}), where the pygmy-resonance strength $\sigma_{\rm py}$, width $\Gamma_{\rm py}$ and centroid $E_{\rm py}$ have been fitted in order to adjust the total theoretical strength function
\begin{equation}
f=K(f_{E1} + f_{M1})+f_{\mathrm{py}} \end{equation}
to the experimental data. The resulting theoretical $\gamma$-strength functions are shown as solid lines in Fig.~\ref{fig:teori}. The obtained values of the fitting parameters for the pygmy resonance are  
together with the normalization constant $K$ 
given in Table~\ref{tab:fit}. The temperature parameter was first kept as a free parameter in the fit, giving $T \approx$ 0.5 MeV for $^{149}$Sm, however the fit did not converge in the case of $^{148}$Sm. Therefore, in order to be consistent we fixed the temperature parameter at 0.5 MeV in both cases and a more pronounced $\chi^2$ minimum was obtained. The chosen temperature of $T =  0.5$ MeV is in good agreement with the average temperature found in the micro canonical ensemble as shown in Fig.~\ref{fig:fig2}.
However, as it has been discussed in \cite{voinov} the temperature parameter in (16) may not correspond to the real nuclear temperature, due to possible uncertainties in the temperature
dependence of the GEDR width.

Figure~\ref{fig:teori} also shows the predicted individual contributions from the giant electric dipole resonance $f_{E1}$, the giant magnetic dipole resonance $f_{M1}$ and the pygmy resonance $f_{\rm py}$ to the total $\gamma$-strength function. The strength function is generally dominated by $E1$ radiation. The $M1$-strength function $f_{M1}$ is $\sim 20$\% of the $E1$-strength function $f_{E1}$ and the pygmy resonance contributes considerably around $3$ MeV.

\begin{figure}[htb]
  \centering
  \includegraphics[angle=-0,width=5.5in]
                {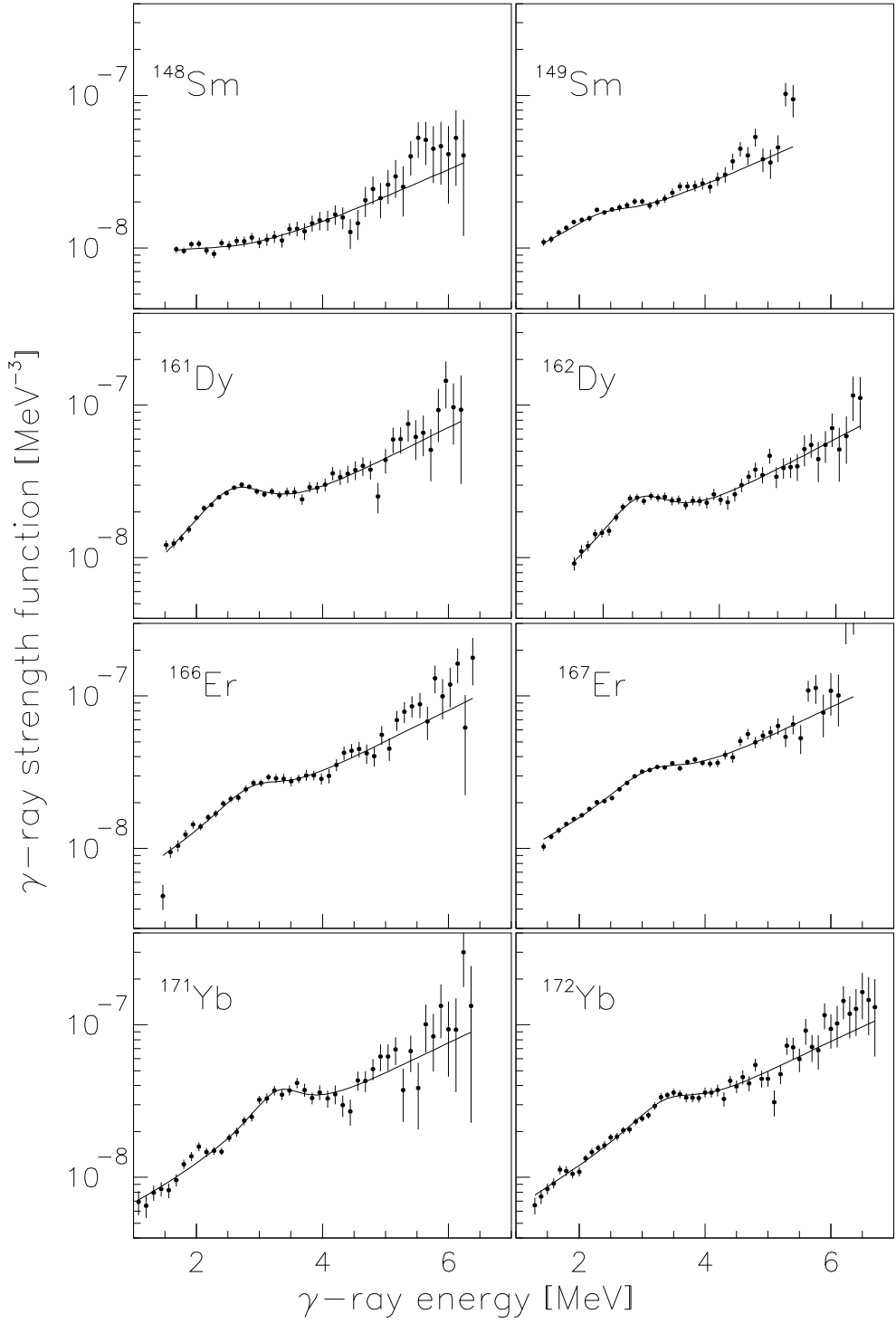}
\caption{The $\gamma$-strength functions for $^{148,149}$Sm, $^{161,162}$Dy, $^{166,167}$Er and $^{171,172}$Yb. The filled circles are experimental data. The solid lines are a fit to the data using the KMF GEDR model and a Lorentzian spin flip GMDR model, and a pygmy resonance.}
\label{fig:strength}
\end{figure}

The $\gamma$-strength functions of $^{148}$Sm and $^{149}$Sm are compared with the $\gamma$-strength functions of the well deformed $^{161,162}$Dy\cite{voinov}, $^{166,167}$Er\cite{elin01} and $^{171,172}$Yb\cite{voinov} nuclei in Fig.~\ref{fig:strength}. The $\gamma$-strength functions of $^{148}$Sm is less than for $^{149}$Sm.  
Some of the explanation can be that in the transitional region from deformed to spherical shapes, one neutron can make a significant difference for the nuclear charge distribution.

The pygmy resonance in the $\gamma$-strength functions was first explained in 
Ref.~\cite{lane71} by the enhancement of the $E1$ $\gamma$-strength function.
This interpretation was adopted in Ref.~\cite{voinov} for the pygmy resonance found at $\sim$ 3 MeV in $^{161,162}$Dy and $^{171,172}$Yb. According to systematic investigations of the pygmy resonance parameters in the rare earth nuclei \cite{iga} it should appear around 1.7 MeV for $^{148,149}$Sm. However, the possibility that the pygmy resonance is of $M1$ character can not be excluded: at an excitation energy around $3$ MeV, there is a concentration of orbital $M1$ strength in the weakly collective scissors mode~\cite{rich}. The scissors mode was first observed in electron-scattering experiments~\cite{riht}, and is confirmed by the $(\gamma,\gamma ')$ reaction~\cite{pietralla}. Recent investigations of Dy and Yb isotopes \cite{scissor} indicate that the observed bump can be interpreted as due to the scissors mode built on exited states. 
The 
present $\gamma$-strength functions show a bump around 2.5 MeV for $^{149}$Sm and around 2.0~MeV for $^{148}$Sm, as shown in Fig.~\ref{fig:teori}. 
The pygmy resonance in $^{148}$Sm $\gamma$-strength function is less obvious. It might be that shell effects in $^{148}$Sm hide the fine structure in the $\gamma$-strength function. Also in the (n,$\gamma$)-data for $^{148}$Sm in Fig.~\ref{fig:iga} it is difficult to locate a $\gamma$-ray bump in the $\sim$ 3 MeV region.

In Ref.~\cite{pietralla2} the total $M$1 strength, observed experimentally in $(\gamma,\gamma ')$ reactions, is found to be: $\sum B(M1)\uparrow = 0.43\mu_N^2$ for $^{148}$Sm. Our data can be compared to this result using
\begin{equation}
\sum B(M1)\uparrow = \frac {\sigma_{\rm py}\Gamma_{\rm py}}{E_{\rm py}}\frac{9\hbar c}{32 \pi^2}.
\end{equation}
For details see Ref.~\cite{scissor}. We obtain 3.5~$\mu_N^2$ for $^{148}$Sm and 2.2~$\mu_N^2$ for $^{149}$Sm. The difference of a factor 10 between the results for $^{148}$Sm can be understood by the fact that in Ref.~\cite{pietralla2} only $\gamma$-rays between 2.7~MeV and 3.7~MeV are taken into account. They are therefore only summing up a part of the high energy tail of the resonance. In addition weak $\gamma$-lines probably have escaped detection, as discussed in Ref.~\cite{scissor}.

\begin{figure}[htb]
  \centering
  \includegraphics[angle=-0,width=5.0in]
                {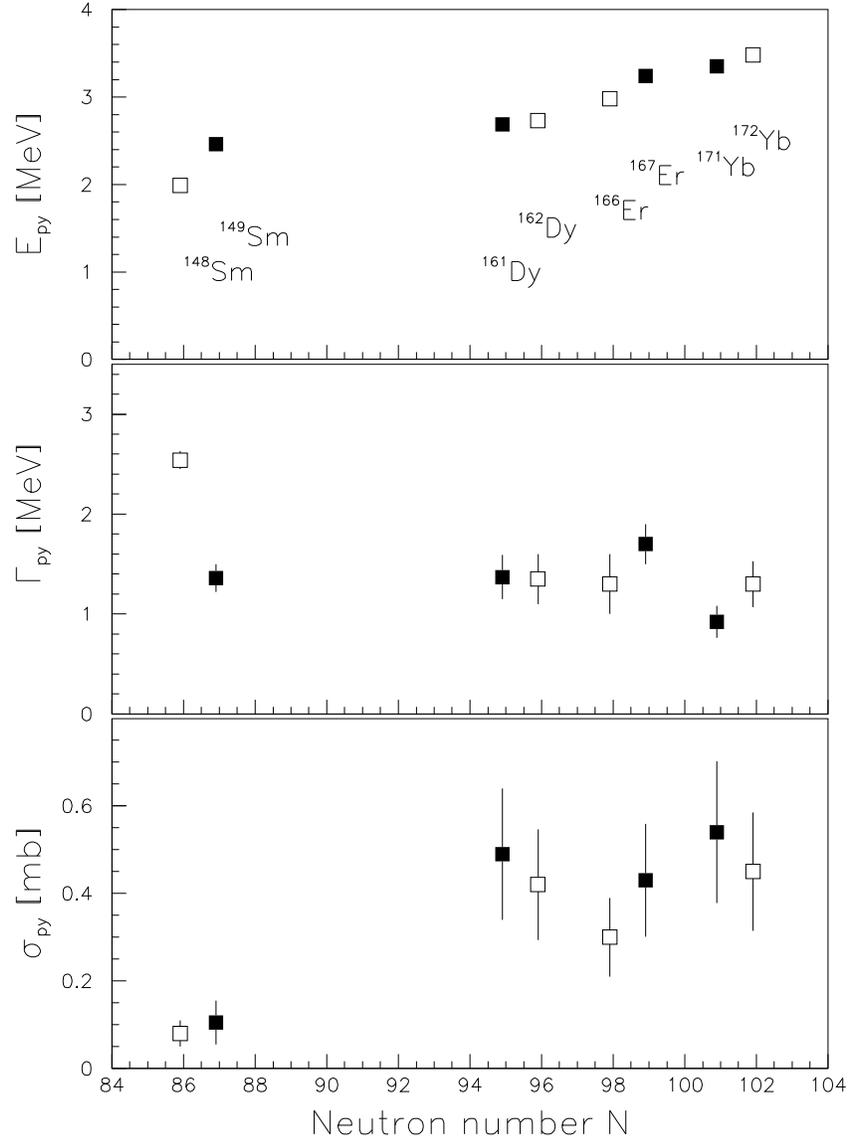}
\caption{Systematics of the pygmy resonance parameters for odd (filled squares) and even (open squares) nuclei as a function of neutron number $N$. The resonance energy $E_{\mathrm{py}}$, the width $\Gamma_{\mathrm{py}}$  and the cross sections $\sigma_{\mathrm{py}}$ with error bars are shown in the upper, middle and lower panels, respectively.}
\label{fig:compar}
\end{figure}

The pygmy resonance parameters are shown in Fig.~\ref{fig:compar} and we believe that the pygmy resonance in the samarium data is of the same nature as the resonance observed in the other rare earth nuclei. Looking at Fig.~\ref{fig:compar} there seems to be a small odd-even effect in the strength of the pygmy resonance $\sigma_{\mathrm{py}}$.
Clarification of the multipolarity of this pygmy resonance awaits new experimental data.

\begin{figure}[htb]
  \centering
  \includegraphics[angle=-0,width=7.in]
                {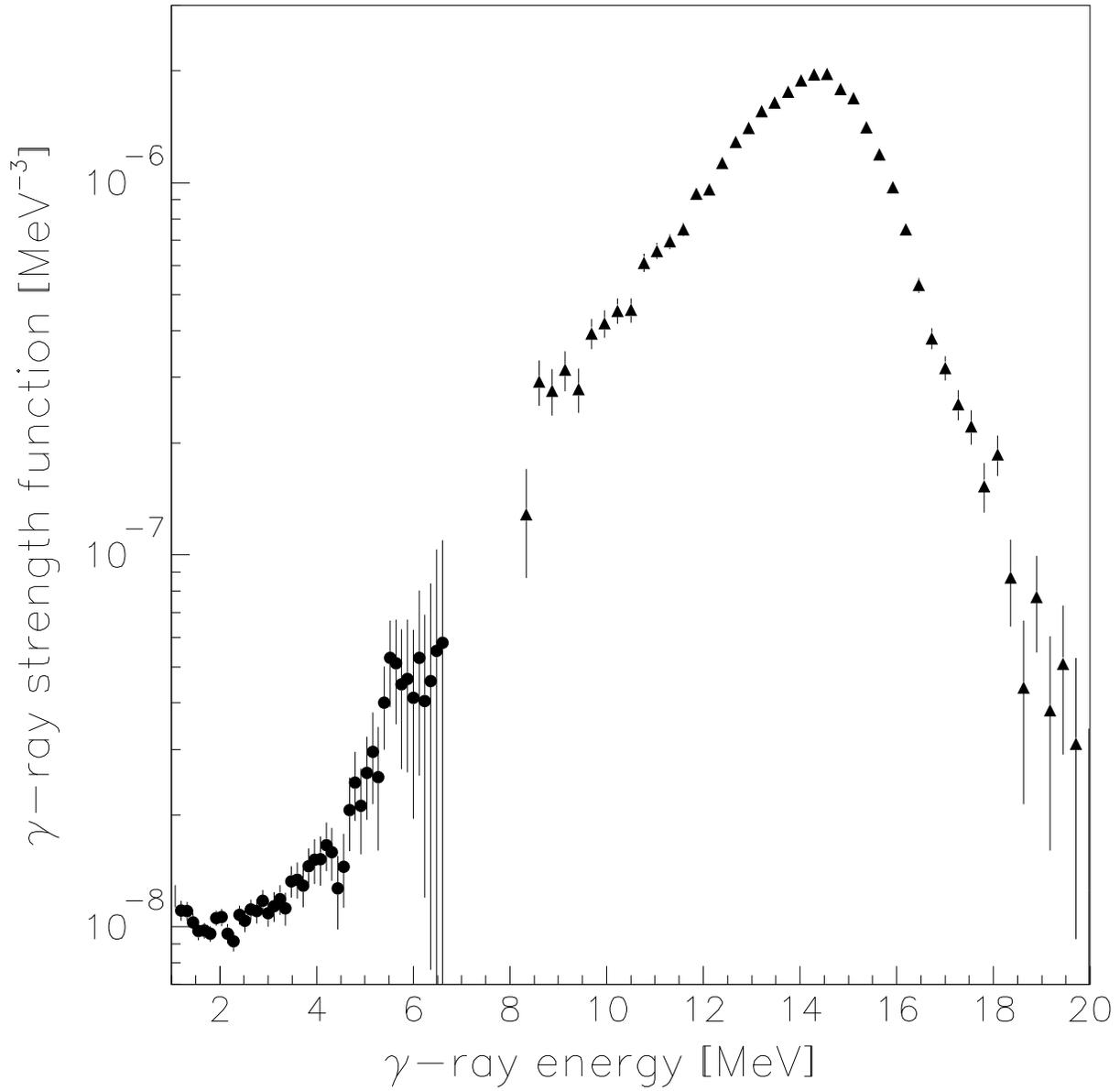}
\caption{The experimental $\gamma$-strength function for $^{148}$Sm ($f_{E1}+f_{M1}$) from the present ($^3$He,$\alpha$)-reaction (circles) plotted together with data obtained from photo-absorption cross section Ref.~\protect\cite{GEDR} (triangles).}
\label{fig:GDR}
\end{figure}

For the first time a comparison of our data to the GEDR data \cite{GEDR} is performed, see Fig.~\ref{fig:GDR}. The tail of the GEDR was until now not very well known and has been modeled in several ways. Our experimental data represent a unique guideline to theoretically describe the shape of this tail.

\section{Conclusions}
The level density and the $\gamma$-strength function have been extracted experimentally for $^{148}$Sm and $^{149}$Sm. The level density of the samarium nuclei is smaller than for the mid-shell dysprosium, erbium and ytterbium nuclei investigated earlier.  From the level densities thermo-dynamical quantities such as temperature and heat capacity were found. Structures in the micro-canonical temperature are interpreted as the onset of new degrees of freedom by the  breaking of Cooper pairs. The S-shape in the heat capacity curves, found within the canonical ensemble, indicates the pairing-phase transition, and a critical temperature for the quenching of pair correlations is found. The S-shape in the heat capacity curve is slightly more pronounced in the odd $^{149}$Sm nuclei, the opposite of what has been observed in the well deformed nuclei, where the even isotopes exhibit more pronounced S-shapes than the odd isotopes.

For the total $\gamma$-spectra of the $^{148}$Sm and 
$^{149}$Sm nuclei, from excitation energy at the neutron binding energy, there is very good agreement between the neutron capture data from the $^{147}$Sm(n,$\gamma$)$^{148}$Sm and $^{148}$Sm(n,$\gamma$)$^{149}$Sm reactions and the same spectrum calculated from the $\gamma$-strength functions and level densities extracted from the present $^{148}$Sm($^3$He,$\alpha$)$^{149}$Sm  and $^{148}$Sm($^3$He,$^3$He')$^{148}$Sm data.

The interpretation of the fine structure of the $\gamma$-strength functions is still an open question, since there are no experimental data to determine if the multipolarity of the pygmy resonance is of $M1$ or $E1$ nature. If it turns out to be $M1$, the resonance at $\sim$~3~MeV in the $\gamma$-strength function 
is likely to be interpreted as due to the scissors mode built on exited states.  

For the first time we compare our data to the (n,$\gamma$) GEDR data and the $^{148}$Sm  $\gamma$-strength function seems to give a consistent extrapolation down to lower $\gamma$-energies. The tail of the GEDR was until now not very well known and has been modeled in several ways, and our experimental data will assist the theoretical effort to describe the shape of this tail.

\section*{Acknowledgments}
The authors would like to thank M. Igashira and S. Mizuno for sending us their neutron capture data from the $^{147}$Sm(n,$\gamma$)$^{148}$Sm and $^{148}$Sm(n,$\gamma$)$^{149}$Sm reactions. The authors also wish to thank Jette S{\"o}rensen for making the target and E.A.~Olsen and J.~Wikne for excellent experimental conditions. Financial support from the Norwegian Research Council (NFR) is gratefully acknowledged.

\end{multicols}

\newpage

\begin{table}
\caption{Parameters used in the normalization of the level 
densities and $\gamma$-strength functions.  
The values of the different parameters are taken from  $^1$ Ref.~\protect\cite{tecdoc},
$^3$ Ref.\protect\cite{GC65}
$^4$ Ref.\protect\cite{IM92} and $^5$ Ref.\protect\cite{NPA93}.} 
\begin{tabular}{ccccccc}
Nucleus    & $\rho$(B$_n$)    & B$_n$ & a         & D      &$f$    &$\langle\Gamma\rangle$ \\
           &(10$^6$MeV$^{-1}$)&(MeV)  &(MeV$^{-1}$)& (eV)   &       & (meV) \\ \hline         \\
$^{148}$Sm &   1.59(18)       & 8.14  & 19.18$^3$ & 5.7(5)$^4$ & 1.1 & 69(2)$^1$ \\
$^{149}$Sm &   0.49(11)       & 5.87  & 19.85$^3$ & 100(20)$^1$& 1.6 & 45(2)$^5$  \\
\end{tabular}
\label{tab:par}
\end{table}

\begin{table}
\caption{Parameters obtained from the pygmy resonance in the $\gamma$-ray strength function. The fitting procedure is performed with the temperature as a fixed parameter of $T = 0.5$ MeV.} 
\begin{tabular}{ccccc}
Nucleus&$E_{\mathrm{py}}$&$\sigma_{\mathrm{py}}$&$\Gamma_{\mathrm{py}}$ & K\\
          &(MeV)    &(mb)       &(MeV)       & \\ \hline
$^{148}$Sm& 1.99(8) & 0.08(3)   & 2.5(2)     & 0.61(6)  \\
$^{149}$Sm& 2.46(5) & 0.11(5)   & 1.4(2)     & 1.1(3)\\
\end{tabular}
\label{tab:fit}
\end{table}

\newpage

\newpage

\end{document}